\documentclass[preprint,aps,showpacs]{revtex4}
\usepackage{amsmath}
\usepackage{amssymb}
\usepackage{bbold}
\usepackage{mathrsfs}
\usepackage{graphicx, psfrag}
\usepackage{amssymb}
\usepackage[colorlinks=true, citecolor=blue, linkcolor= red, bookmarks=true]{hyperref}
\usepackage{xr}
\usepackage{zref-xr}

\begin{document}

\def \beq{\begin{equation}}
\def \eeq{\end{equation}}
\def \bse{\begin{subequations}}
\def \ese{\end{subequations}}
\def \bea{\begin{eqnarray}}
\def \eea{\end{eqnarray}}
\def \bem{\begin{displaymath}}
\def \eem{\end{displaymath}}
\def \bem{\begin{bmatrix}}
\def \eem{\end{bmatrix}}
\def \Ps{\hat{\Psi}(\boldsymbol{r})}
\def \Pds{\hat{\Psi}^{\dagger}(\boldsymbol{r})}
\def \i{{\int}d^2{\bf r}}
\def \bl{\bar{\boldsymbol{l}}}
\def \c{\hat{c}_{n,m}}
\def \cp{\hat{c}_{n',m'}}
\def \cd{\hat{c}_{n,m}^{\dagger}}
\def \cdp{\hat{c}_{n',m'}^{\dagger}}
\def \bb{\bibitem}
\def \nn{\nonumber}
\def \bs{\boldsymbol}

\title{\textbf{Cavity optomechanics with synthetic Landau levels of ultra cold Fermi gas }}
\author{ Bikash Padhi and Sankalpa Ghosh}
\affiliation{Department of Physics, Indian Institute of Technology Delhi, New Delhi-110016, India}
\email {sankalpa@physics.iitd.ac.in}
\begin{abstract}
Ultra cold fermionic atoms placed in a synthetic magnetic field arrange themselves in Landau levels. We theoretically study the optomechanical interaction between 
the light field and collective excitations of such fermionic atoms  in synthetic magnetic field by placing them in side a Fabry Perot cavity. We derive 
the effective hamiltonian for particle hole excitations from a filled Landau level using a bosonization technique and obtain an expression for the cavity transmission 
spectrum. Using this we show that the cavity transmission spectrum demonstrates  cold atom analogue of Subnikov de Hass oscillation in electronic condensed matter systems. 
We discuss the experimental consequences for this oscillation for such system and the related optical bistability. 
\end{abstract}

\maketitle

By allowing an ultracold atomic ensemble to interact with a selected mode of a high finesse cavity
it is possible to probe the quantum many body state of such an ultra cold atomic system. The resulting   
cavity-optomechanics or cavity quantum electrodynamics with  ultra cold atoms in recent times
witnessed significant development \cite{review1}. The experimental successes include  the coupling of collective density excitation of 
a ultra cold bosonic condensate with a single cavity mode and observing the coupled dynamics through cavity transmission \cite{Eslng},
a strongly  coupled cavity mode with a highly localized ultra cold atomic condensate trapped inside a single antinode of a cavity field \cite{Colombe},
 demonstration of strong ultracold atom-cavity coupling induced optical non-linearity even at low photon density \cite{Kurn} and selected atom-photn coupling 
of single atomic ensemble in a multi-ensemble system \cite{Kurn1} to name a few. 

In another development,  there has been significant experimental and theoretical progress in studying the effect of 
optically induced artificial or synthetic  gauge field \cite{artificial} on such neutral atoms, making it a playground for quantum simulation of phenomena that   
occur when electronic condensed matter system is placed in a real magnetic field. The experimental achievements in this direction 
include the observation of vortices, Abrikosov vortex lattice in trapped ultra cold atomic superfluid initially by achieving 
the synthetic magnetic field  through the rotation of the trap\cite{vortices}, and later by Raman laser induced spatially varying 
coupling of the hyperfine states of such ultra cold atoms \cite{synthetic}. More recent development in this direction includes 
the creation of optical flux lattices \cite{opticalflux}, realization of 
spin-orbit coupling for such neutral ultracold bosonic \cite{spinorbitB} and fermionic atoms \cite{spinorbitF} with 
the possibility of creating ultra cold atomic analogues of topological condensed matter phases.
 
\begin{figure}
\begin{center}
\includegraphics[scale=0.3]{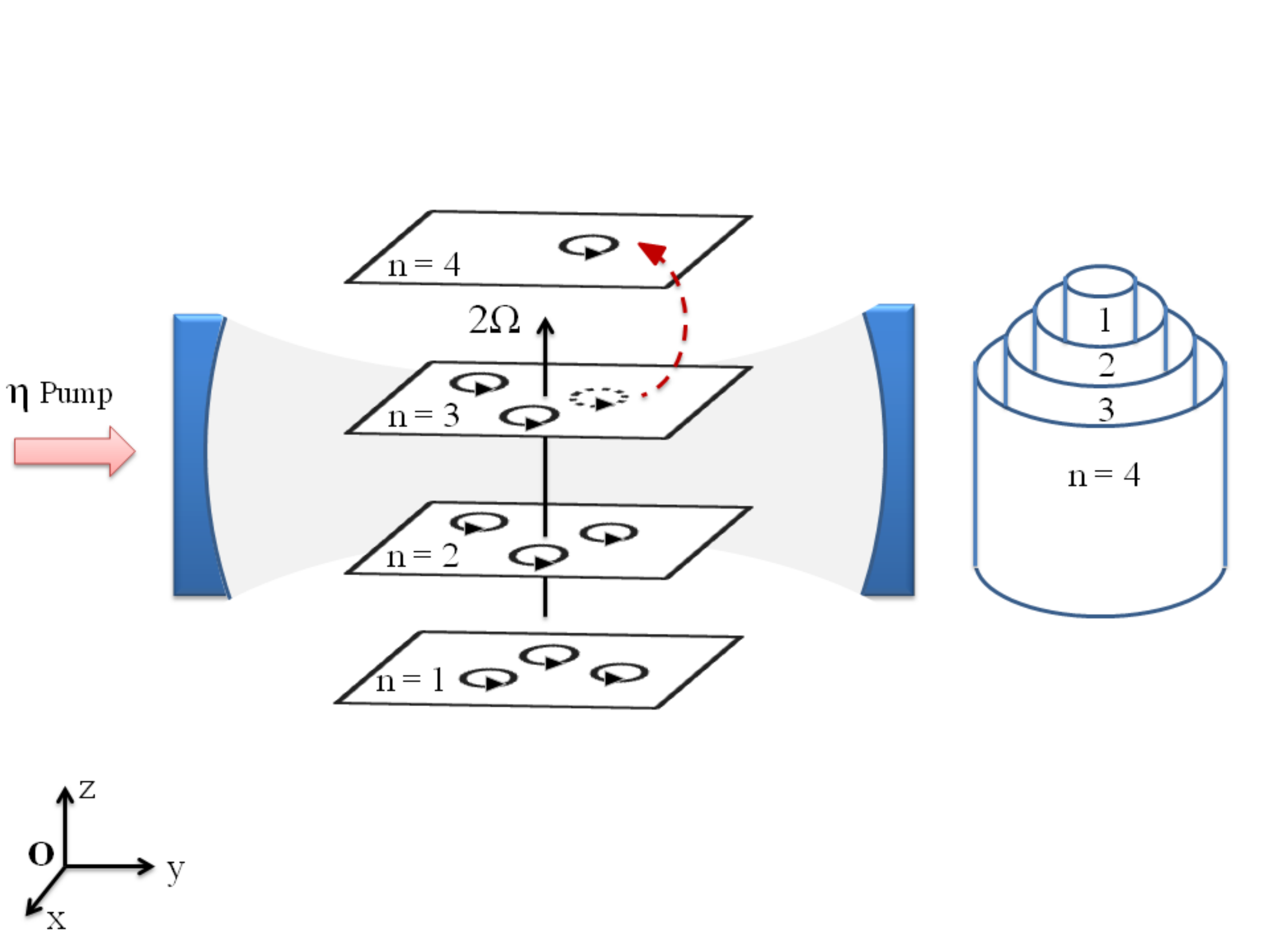}
\caption{\small \sl Schematic diagram of the system considered is shown. The concentric cylindrical surfaces represent 
the fermi surfaces that correspond to  Landau levels of (LL)  of fermionic atoms inside the cavity with LL quantum number 
$n$. The particle 
hole excitation from the last filled Landau level is also also shown.  
 \label{setup}}
\end{center}
\end{figure}

This letter aims to combine  these two developments by 
considering  a system of such ultra cold atoms trapped inside a high finesse Fabry Perot cavity  interacting with 
a single cavity mode ( see Fig. \ref{setup} ), additionally, in the presence of a synthetic magnetic field. Specifically  we consider the case of 
ultra cold fermionic atoms \cite{Jin1} in a synthetic magnetic field \cite{Ho, Lewen}
such that a set of Landau levels can be filled according to Pauli principle. We consider the coupling between the bosonic 
particle-hole like excitations from such filled Landau levels of  fermionic atoms with the cavity mode. We find that the atom-photon coupling 
explicitly shows Landau level degeneracy, and has  
finite discontinuities at certain values of artificial magnetic field strength that resembles the well known Subnokov de Haas oscillation \cite{SdH} in condensed matter system.
The cavity transmission spectrum shows optical bistability, a hall mark of optical non-linearity in such cavity system \cite{Meystre}, but now the features of the bistable curve also reveals 
the Landau level structure. Our results suggest that cavity optomechanics with such atomic Landau levels can be a powerful probe for ultra cold atoms in 
synthetic gauge field.

We consider two dimensional system of N ultracold neutral fermionic two-level atoms 
each of mass $M$ subjected to a synthetic magnetic field \cite{synthetic, Ho, Lewen, review}, placed inside  Fabry-P${\acute e}$rot cavity of area $\mathcal{A}$ which is 
driven at the rate of ${\eta}$ by a pump laser of frequency ${\omega}_p$ and wave vector $\bs{K} = (K_x, K_y)$. 
%$K_x = K_y = |{\bf K}|/\sqrt{2}$. 
The atoms have transition frequency $\omega_a$, and interact strongly with a single standing wave empty cavity mode of frequency $\omega_c$. 
%%%%%%%%%%%%%
%%%%%%%%%%%%%
We take the artificial magnetic field as $2 \Omega \hat{z}$. 
%If the synthetic gauge field is created by rotating the trap potential in the $x-y$ plane, then $\Omega$ can identified 
%with the rotational frequency. 
We also ignore the effective trap potential assuming it is shallow enough in the bulk. The resulting single particle hamiltonian 
is analogous to Landau problem of a charged particle in a transverse magnetic field  ( for details see \cite{suppli1}) that can be 
written as.
\beq
H_L=\frac{1}{2M}{\bf \Pi}^2,	\label{HLE}
\eeq
where ${\bf \Pi}={\bf p} - M{\bf A}$ is the kinetic momentum, with the effective vector potential 
$ {\bf A} = {\bf \Omega \times r}$ in symmetric gauge. 
The eigenstates of this Hamiltonian are Landau levels with effective cyclotron frequency $\omega_0 = 2\Omega$, and eigen-energies $E_{n,m}  = 2\hbar\Omega(n + 1/2)$. The effective magnetic length in this problem is $l_0 = \sqrt{\hbar/2M\Omega}$. 

If the pump laser frequency $\omega_p$ is far detuned from the atomic transition frequency $\omega_a$, the excited electronic state of the two level 
atoms can be adiabatically eliminated. It is assumed that such atoms interact dispersively with the cavity field, taken to be single mode. In the dipole  
and rotating wave approximation, we get the effective system Hamiltonian( details in \cite{suppli1})

%\beq
%\hat{H}_{eff} = {\i}{\Pds}\Big{[}{\bf \hat{\Pi}}^2/2M + \hbar U_0cos^2({\bf K}.{\bf r})\hat{a}^\dagger \hat{a}\Big{]}{\Ps} + {\hbar{\Delta_c}}\hat{a}^{\dagger}\hat{a}-{\imath}{\hbar}{\eta}%(\hat{a}-\hat{a}^\dagger). \label{hfull}\eeq
\bea
\hat{H}_{eff} & = & \hat{H}_L + \hat{H}_I + \hat{H}_C , \text{with} \label{hfull} \\
\hat{H}_L&= & {\i}{\Pds}\Big{[}{\bf \hat{\Pi}}^2/2M \Big{]}{\Ps},  \label{HL} \\
\hat{H}_I&=  & {\i}{\Pds}\Big{[}\hbar U_0cos^2({\bf K}.{\bf r})\hat{a}^\dagger \hat{a}\Big{]}{\Ps}, \label{HI} \\
\hat{H}_C&=  & {\hbar{\Delta_c}}\hat{a}^{\dagger}\hat{a}-{\imath}{\hbar}{\eta}(\hat{a}-\hat{a}^\dagger). \label{HC}
\eea
Here $U_0 =g_0^2/\Delta_a$ is the effective light-matter coupling constant with $g_0$ is single photon Rabi frequency, $\Delta_{a} = \omega_p-\omega_a$. Here $\hat{H}_L$ is the atomic Hamiltonian in the synthetic magnetic field, $\hat{H}_C$ captures the dynamics of the cavity photons with $\Delta_{c}=\omega_{c} - \omega_{p}$. 
$\hat{H}_I$ is the term that describes interaction between atom and the cavity mode.  
%\bse \begin{align}
%\hat{H}_L&= {\i}{\Pds}\Big{[}{\bf \hat{\Pi}}^2/2M \Big{]}{\Ps},  \label{HL} \\
%\hat{H}_I&= {\i}{\Pds}\Big{[}\hbar U_0cos^2({\bf K}.{\bf r})\hat{a}^\dagger \hat{a}\Big{]}{\Ps}, \label{HI} \\
%\hat{H}_C&= {\hbar{\Delta_c}}\hat{a}^{\dagger}\hat{a}-{\imath}{\hbar}{\eta}(\hat{a}-\hat{a}^\dagger). \label{HC}
%\end{align} \ese
The  atomic field operator in the Landau level basis (symmetric gauge) is given as 
\bea \Ps & = &\sum\limits_{m,n}\c \langle{\bf r}|n,m\rangle
= \sum\limits_{m,n}\frac{e^{-|z|^2/4l_0^2}}{\sqrt{2\pi l_0^2}}G_{m+n,n}(iz/l_0) \c   \label{symmetric} \\
\text{with}, \{\cd,\hat{c}^\dag_{n',m'}\} &=& \{\c,\hat{c}_{n',m'}\} = 0  \nn \\
\{\hat{c}^\dag_{n,m},\hat{c}_{n',m'}\} &=& \delta_{n,n'}\delta_{m,m'},	\label{cnmcommut}
\eea
%\bea
%\Ps& = \sum\limits_{m,n}\c \langle{\bf r}|n,m\rangle
%& = \sum\limits_{m,n}\frac{e^{-|z|^2/4l_0^2}}{\sqrt{2\pi l_0^2}}G_{m+n,n}(iz/l_0) \c , \nn \\
%\Pds& = \sum\limits_{m,n}\cd \langle n,m|{\bf r}\rangle
%& = \sum\limits_{m,n}\frac{e^{-|z|^2/4l_0^2}}{\sqrt{2\pi l_0^2}}G^*_{m+n,n}(iz/l_0) \cd . \label{expansion}
%\eea
with $z=x+\imath y$. $|n,m\rangle$ is the Landau-eigenket. $\langle{\bf r}|n,m\rangle$ is the symmetric-gauge wavefunction. 
$\frac{e^{-|z|^2/4l_0^2}}{\sqrt{2\pi l_0^2}}G_{n+m,m}(iz/l_{0})$  are two dimensional harmonic oscillator wavefunction 
whose properties are given in \cite{suppli1}.

%\ref{Gmn} the function $G_{n+m,m}(z)$ is formally derived and its properties are discussed in full detail.
$\cd$ is the fermionic creation operator that creates the state $|n,m \rangle$, namely 
a fermion in the $n^{th}$ Landau level (LL), with the guiding center $m$ obeying (\ref{cnmcommut})
with $n=0,1,2...,\nu-1$ and $m = 0,1,2,..., N_\phi -1$.  $\nu= N/N_\phi$ is called the filling factor where $N_\phi=\mathcal{A}/(2\pi l_0^2)$ is the degeneracy of each Landau level. 
The atomic  Hamiltonian ($\hat{H}_L$) can be diagonalized in the Landau level basis yielding 
$\hat{H}_L=\hbar {\omega}_0 \sum\limits_{m,n=0}^{\infty} (n+\frac {1}{2})\hat{c}_{n,m}^{\dagger} \c.$ 
Using $4cos^2({\bf K.r})=2+2cos(2{\bf K.r})= 2+e^{-\imath 2{\bf K.r}}+e^{\imath 2{\bf K.r}}$, $\hat{H}_I$) in the Landau basis can be written as  
\bea
\hat{H}_I&=&\frac{\hbar U_0}{4}\Big{\{} \hat{N} + \sum\limits \cdp \c e^{-2(|K|l_0)^2} \Big (G_{n',n}(2 K^*l_0)G_{m',m}(2Kl_0) + \nonumber \\ 
& & G_{n',n}(-2K^*l_0)G_{m',m}(-2Kl_0)\Big )\Big{\}}\hat{a}^\dagger \hat{a}  \label{HFermi}
\eea
where $ K =  K_x + \imath K_y $, $K^{*}=K_x -\imath K_y$, $ |K|^2 =  K_x^2 +  K_y^2$, and the summations are done over all available $n, n',m,m'$. If we assume that the interaction 
time between the cavity and ultra cold fermions is much shorter than any time scale associated with the reorganization of the atomic ground state in presence 
of the standing wave inside the cavity, the role of  the interaction hamiltonian (\ref{HFermi})  is restricted  to  transfer momentum $\pm 2|K|$ to the particle-hole excitation above the 
Fermi level. 

%%%%%%%%%%%%%%%%%%%%%%%%%%%%%%%%%%

\def \cp{\hat{c}_{n+p,m'}}
\def \cdp{\hat{c}_{n+p,m'}^\dag}

By looking at the cavity transmission spectrum we are interested in studying such low energy excitations above an integer number of filled Landau level. 
Such particle-hole excitations are bosonic in nature and  known as 
magnetic exciton in the literature of quantum hall systems \cite{Halperin}.
In the absence of atom-photon interaction 
the ground state of our system is a direct product state of photonic vacuum and excitonic vacuum,  obtained by completely filling the first $\nu$ Landau levels of each guiding center 
\beq
|GS\rangle = \prod \limits_{m=0}^{N_\phi-1} \prod \limits_{n=0}^{\nu-1} c_{n,m}^{\dagger} |0\rangle. \label{GS}
\eeq 
These inter Landau level excitations only involve the change in the 
Landau level index, they can be studied using the language of bosonization \cite{Calderia} by introducing  
bosonic operator 
\beq
\hat{b}^\dag_p({\bf k})=\frac{1}{\sqrt{pN_\phi J_p^2(k R_\nu)}}
e^{-(l_0|k|)^2/2} \sum\limits_{n=0}^{\infty} \sum\limits_{m,m'=0}^{\infty} \cdp \c \Big (G_{n+p,n}(l_0 k^*)G_{m',m}(l_0k)\Big ). 
\label{boperator}
\eeq
This operator creates a bosonic particle-hole excitation by shifting an atom from $n$-th LL to the $n+p$-th LL
where  $J_p$ is the Bessel function of first kind , $R_\nu=\sqrt {2\nu}l_0$, and obeys 
\beq
[ \hat{b}_p({\bf k}), \hat{b}_q({\bf k'})] = [ \hat{b}^\dag_p({\bf k}), \hat{b}^\dag_q({\bf k'})] =0  \nn \eeq
\beq
[ \hat{b}_p({\bf k}), \hat{b}^\dag_q({\bf k'})] = \delta({\bf k}-{\bf k'})\delta_{p,q}.  \label{bcom} \eeq
%%%%%%%%%%

Using the commutators of the bosonic opertaor (\ref{bcom}) 
%$[\hat{H}_L, \hat{b}^\dag_p({\bf k})] = \hbar \omega_0 p \hat{b}^\dag_p({\bf k}) ; [\hat{H}_L, \hat{b}_p({\bf k})] = \hbar \omega_0 p \hat{b}_p({\bf k}) $.  
the bosonized version of the Landau level hamiltonian (\ref{HL}) of ultra cold fermions can be written as  
\beq 
\hat{H}_L= \hbar \sum\limits_{p=1}^{\infty} \sum\limits_{\bf k} p\omega_0  b_p^\dagger({\bf k})b_p({\bf k}). \label{BL} \eeq
The  atom-photon interaction (\ref{HFermi}), can similarly be rewritten in terms of the bosonic operator 
(\ref{boperator}) as 
\beq
\hat{H}_I = \frac{\hbar U_0}{4}\hat{N}\hat{a}^\dagger \hat{a} + \frac{\hbar U_0}{4}\sum\limits_{p=1}^{\infty} \sqrt{N_\phi pJ_p^2(2K R_\nu)}\Big (\hat{b}_p^\dagger(2{\bf K}) +\hat{b}_p(2{\bf K})\Big ) \hat{a}^\dagger \hat{a}.
\label{BAP} \eeq
The derivation of the hamiltonian (\ref{BL}), and (\ref{BAP}) from (\ref{HFermi}) is given in the supplementary information \cite{suppli1}. The  bosonized effective Hamiltonian (\ref{hfull}) of the atom-photon system thus becomes
\bea
\hat{H}_{eff} &=& \hbar \sum\limits_{p=1}^{\infty} \Big \{\sum\limits_{\bf k} p\omega_0 \hat{b}_p^\dagger({\bf k})\hat{b}_p({\bf k}) +  \delta^\nu_p \sqrt{p}\Big (\hat{b}_p^\dagger(2{\bf K}) +\hat{b}_p(2{\bf K})\Big ) \hat{a}^\dagger \hat{a} \Big \} \nn \\ 
&& + {\hbar{\Delta}}\hat{a}^{\dagger}\hat{a} -{\imath}{\hbar}{\eta}(\hat{a}-\hat{a}^\dagger), \label{hboson}\\
\delta^\nu_p&=& \frac{U_0}{4} \sqrt{N_\phi J_p^2(2K R_\nu)}. \eea 
The above hamiltonian is  one of the central result of this paper. 
Here the operator $\hat{N}$ is replaced with its steady-state expectation value, subsequently the term $\frac{N\hbar U_0}{2}\hat{a}^\dagger a$ is  incorporated into ${\hbar{\Delta_c}}a^{\dagger}a$ of $H_C$ to get the effective cavity detuning $\Delta = \omega_c - \omega_p + \frac{NU_0}{2}$. 

\begin{figure}
\centering
\includegraphics[width = 8cm, height = 6cm]{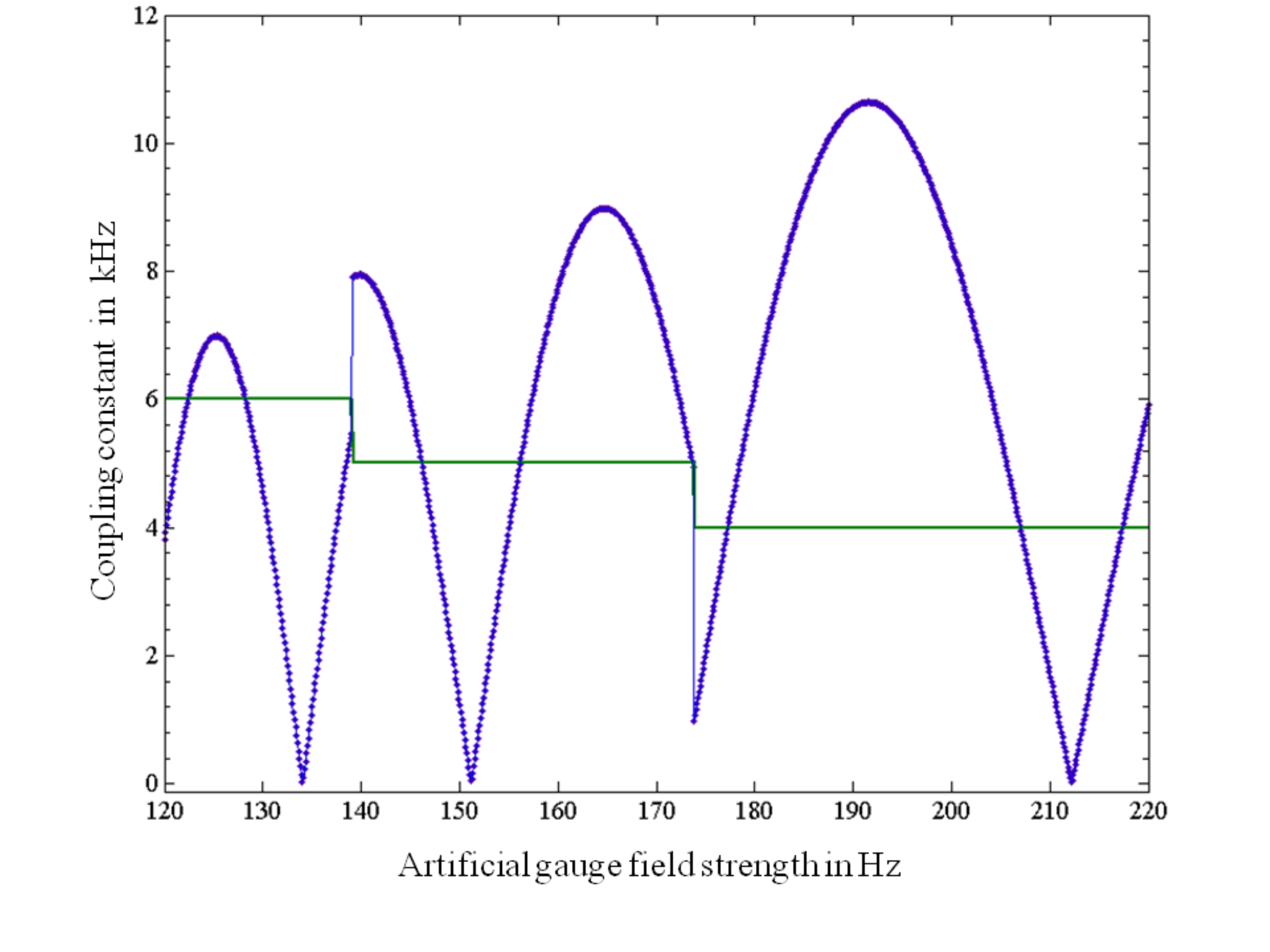}
\caption{\small Variation of coupling constant (for p=1) with synthetic field strength. The green colored steps in the background correspond to the corresponding first empty LL, $\nu$ = 6,5,4.}.
\label{delp_B}
\end{figure}

The $\delta^\nu_p$ is the atom-photon coupling constant that couples the excited  levels with the photon field. 
Fig. (\ref{delp_B}) depicts its variation with the field strength when an atom gets excited from the filled LL 
to the next unoccupied LL. For that purpose we choose 
experimentally achievable parameters, $\lambda = 500nm (K \simeq 10^7m^{-1}), \mathcal{A} \simeq (30\mu m)^2, \kappa = 2\pi MHz$, atomic mass $M = 1.5 \times 10^{-25} kg, N = 2000, g_0 = 2\pi \times 10 MHz,$ pump-atom detuning $\omega_p - \omega_a = 2\pi \times 50 GHz$. 
It lineary depends on  atom-photon coupling constant $U_0$ and is enhanced by the Landau level degenracy ${\sqrt{N_\phi}}$, different as 
compared to the case of ordinary fermions \cite{Meystre} and  akin to the scaling of 
the atom-photon coupling constant by $\sqrt{N}$ for a N-boson condensate \cite{Eslng, Colombe}.    
In Fig. (\ref{delp_B}), 
with increasing $\Omega$, the coupling constant oscillates along with jump discontinuities. This is the usual Subnikov de Hass effect \cite{SdH}, now occuring  
for a synthetic magnetic field  when the fermi level makes a jump to the previous level at some increased value of the field. 

The other important feature, the oscillatory behaviour of $\delta^{\nu}_p$ can be attributed to the length scales associated with the current problem. 
In presence of synthetic gauge field the cyclotron radius of the ultracold atoms ($l_0 \sim 200-800 nm$) is comparable with the wavelength of the probing photon ($\lambda \sim 600 nm$). With increase in field strength the cyclotron radius decreases and the number of wavelengths that fits within this radius also changes, 
leading to the oscillatory behavior of the atom-photon coupling strength as a function of the field strength.
In comparison, in the corresponding electronic problem the electron cyclotron radius is much smaller ($l_0 \sim 20 nm$ ) so the incident photon can not actually \textit{see}  the 
individual cyclotron orbit, making such oscillation hard to be observed in electronic LL spectroscopy \cite{spec1, spec2}.
%\beq
%\delta_p^\nu =\frac{U_0}{4}\ \sqrt{\frac{N_\phi}{2\pi K %R_\nu}}cos(2KR_\nu-(2p+1)\frac{\pi}{4}). 
%\eeq				%%%%%%%%%%%%
 		%%%%%%%%%%%%

The oscillation of the coupling constant as a function of the strength of the synthetic gauge field 
can be obtained from the steady state  cavity transmission spectrum, an experimentally measurable quantity. The 
hamiltonian (\ref{hboson}) represents a coupled system   of particle-hole  excitations ( magnetic exciton) and 
photons. The steady state solution of Heisenberg equations for  
operators associated with the dynamics of  exciton and photon yields the cavity transmission spectrum. To this purpose we introduce phase space quadrature variables 
 \beq \hat{X}_L = (\hat{b}_p^\dagger (2{\bf K}) + \hat{b}_p({2\bf K}))/\sqrt2 ; \hat{P}_L = \imath (\hat{b}_p^\dagger(2{\bf K}) - \hat{b}_p(2{\bf K}))/\sqrt2. 
\nn \eeq
%\bea
%\begin{split}
%\hat{X}_L &=& (\hat{b}_p^\dagger (2{\bf K}) + \hat{b}_p({2\bf K}))/\sqrt2, \\
%\end{split} \quad \quad
%\begin{split}
%\hat{P}_L &=& \imath (\hat{b}_p^\dagger(2{\bf K}) - \hat{b}_p(2{\bf K}))/\sqrt2. 
%\end{split}
%\eea
which obey standard commutator 
$[\hat{X}_L,\hat{P}_L] = \imath$.  The resulting Heisenberg equations are 
\bea 
\frac{d\hat{X}_L}{dt} &=& p\omega_0 \hat{P}_L, \nn \quad \quad \frac{d\hat{P}_L}{dt} = -p\omega_0\hat{X}_L - \delta^\nu_p \sqrt{2p}\hat{a}^\dagger \hat{a}, \nn \\
\frac{d\hat{a}}{dt} &=& -\imath \sum\limits_{p=1}^{\infty} \delta^\nu_p \sqrt{2p}\hat{X}_L \hat{a} - \imath\Delta \hat{a} + \eta - \kappa \hat{a} + \sqrt{2\kappa}\hat{a}_{in}.	\label{dynamics}
\eea

Here $\kappa$ is the cavity decay rate and $\hat{a}_{in}$ denotes a Markovian noise operator\cite {noise} with  zero mean, and correlation $\langle \hat{a}_{in}^\dagger(t)\hat{a}_{in}(t')\rangle = 2\kappa\delta(t-t'),$ and $\langle \hat{a}_{in}(t)\hat{a}_{in}(t')\rangle = 0$, so can  be dropped for steady state analysis. The steady state solutions are given by 
\bea
\hat{P}_L^{(s)} = 0, \quad & & 
\hat{X}_L^{(s)} = -\frac{\delta^\nu_p \sqrt{2p}}{p\omega_0}\hat{a}^{\dag(s)}\hat{a}^{(s)}, \quad \nn \\
\hat{a}^{(s)}  & = &  \frac{\eta}{\kappa + \imath  (\Delta - S_\nu \hat{a}^{\dag(s)}\hat{a}^{(s)})}, \text{with} \nn \\
S_\nu & = & \frac{2}{\omega_0}\sum\limits_{p=1}^{\infty} (\delta^\nu_p)^2 = \frac{U_0^2\mathcal{A}M}{32\pi\hbar}(1-J_0^2(2KR_\nu)). \label{Snu} \eea

% $ S_\nu = \sum\limits_{p=1}^{\infty} \frac{2(\delta^\nu_p)^2}{\omega_0},$ which can be simplified using identities of Bessel functions of first kind \cite{Bessel} 
%\beq
%S_\nu = \frac{2}{\omega_0}\sum\limits_{p=1}^{\infty} (\delta^\nu_p)^2 = \frac{U_0^2\mathcal{A}M}{32\pi\hbar}(1-J_0^2(2KR_\nu)).
%\label{Snu}
%\eeq
Therefore steady state inter cavity photon number is  
\beq
\hat{n}_{ph} = \hat{a}^{\dag(s)}\hat{a}^{(s)}= 
\frac{\eta^2}{\kappa^2 + (\Delta - S_\nu \hat{n}_{ph})^2}.
\eeq
The cavity transmission spectrum is  given by its expectation value that follows 
\bea
S_\nu^2n_{ph}^3 - 2 S_\nu \Delta n_{ph}^2 + (\kappa^2+\Delta^2)n_{ph} = \eta^2.	\label{EOS}
\eea
%%%%%%%%%%%%%
\begin{figure}
%\centering
%\begin{subfigure}[b]{.4\textwidth}
%\centering
\includegraphics[width=6cm, height= 6cm]{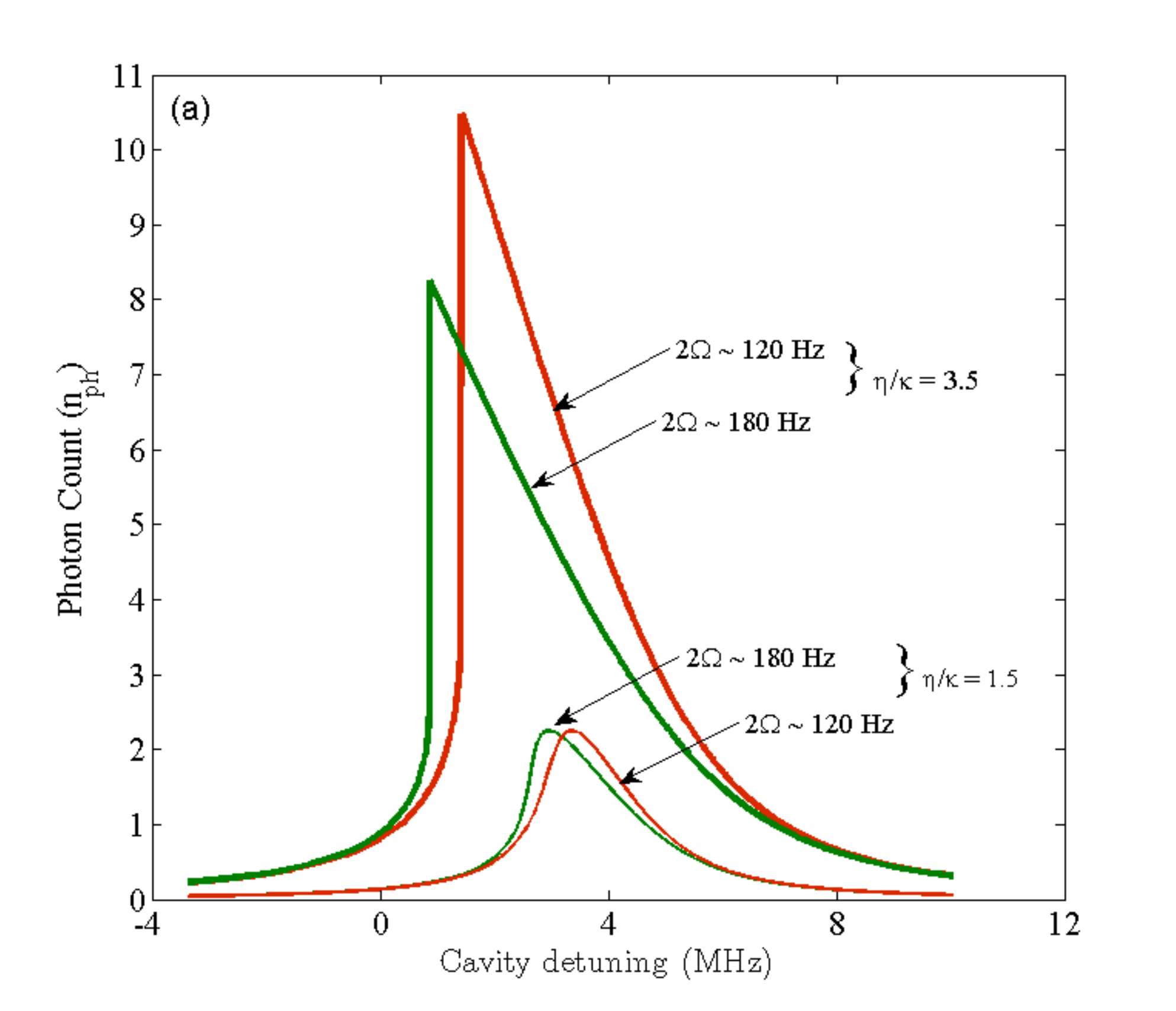}
%\end{subfigure}
%\begin{subfigure}[b]{.4\textwidth}
%\centering
\includegraphics[width=6cm, height= 6cm]{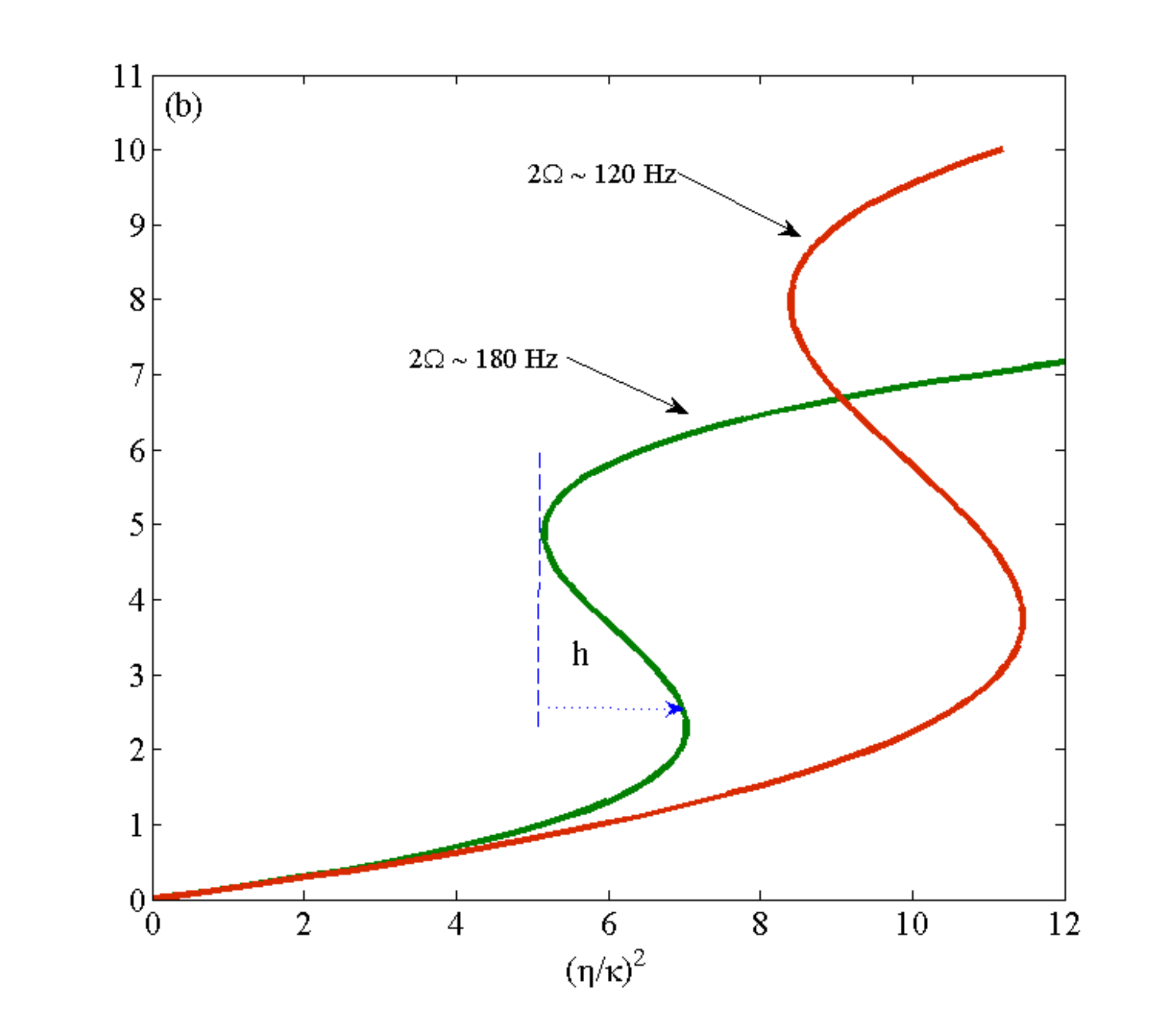}
%\end{subfigure}
\caption{ Steady-state interactivity photon number as a function of (a) pump cavity detuning for a set of synthetic field and $\eta/\kappa$; (b) pump rate for the same set of synthetic fields and cavity detuning $\Delta = 2\pi \times$ 2.5 MHz.}
\label{eta}
\end{figure}
%%%%%%%%%%%%%
Such non-linear cubic equation is characteristic of optical multistability \cite{Eslng, Kurn}. Figure \ref{eta} shows the behavior of steady-state mean photon number as a function of pump rate, and 
$\Delta$.  To understand this multistability we 
study the fluctuation around the steady state, through a  linear stability analysis. 
To that purpose, we write $n_{ph} = n_{ph}^{s} + \delta n_{ph}$ in Eq. (\ref{EOS}) where $n_{ph}^{s}$ corresponds to the steady state inter cavity photon number plotted in Fig. \ref{eta} (b). 
Then only the terms linear in $\delta n_{ph}$ are kept in the resulting equation, and the steady state solution (\ref{EOS}) is substituted in it to get 
\beq  \left(  3S_\nu^2(n_{ph}^{s})^2 - 4 S_\nu \Delta n_{ph}^{s} + (\kappa^2+\Delta^2)\right ) \delta n_{ph} = 0.\label{LSA} \eeq
The solution of this equation
 defines  the upper and lower bound of the unstable regime as the turning points of the plot in Fig. \ref{eta} (b). In the region 
between these two turning points, the  inter cavity photon number is a decreasing function of the cavity parameter $(\frac{\eta}{\kappa})^{2}$ 
and corresponds to  the unstable solution \cite{MeystreB, Gibbs}.   

A more formal way of doing the linear stability 
analysis is through the 
Heisenberg equation of motion of the operators, namely setting  $\mathcal{O}(t) = \mathcal{O}^{(s)} + \delta \mathcal{O}(t)$.
However for the current problem $\mathcal{O}(t) = [\hat{X}_L (t) ^{p=1} , \hat{P}_L (t) ^{p=1} ,  \hat{X}_L(t)^{p=2}, \hat{P}_L(t)^{p=2}, \cdots,  \hat{X} (t) , \hat{P} (t) ]^T,$ 
with $\hat{X} =  (\hat{a}^\dag + \hat{a})/\sqrt2, \hat{P}= \imath(\hat{a}^\dag - \hat{a})/\sqrt2$ being the cavity quadratures. 
The linear stability analysis gets contribution from all $p$~s and the resulting stability matrix is infinite dimensional and cannot be handled in the same 
way like one in the absence of  such  synthetic  gauge field \cite{Meystre}.  A more detailed discussion on this issue is given in the 
supplimentary information \cite{suppli1}. 
%%%%%%%%%%%%%%
\begin{figure}
%\centering
%\begin{subfigure}[b]{.4\textwidth}
%\centering
\includegraphics[width=7.3cm, height= 6cm]{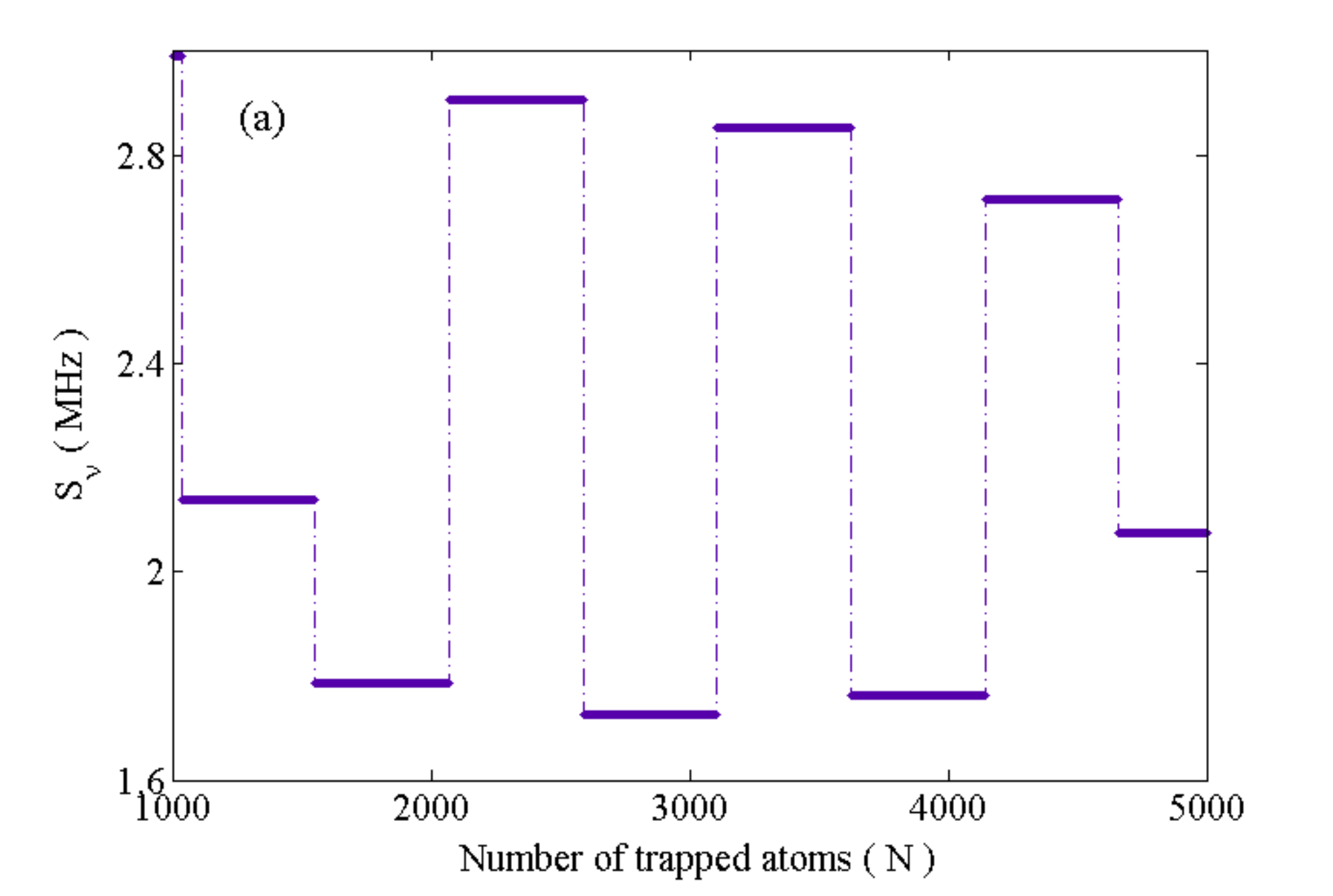}
%\end{subfigure}
%\begin{subfigure}[b]{.4\textwidth}
%\centering
\includegraphics[width=7.3cm, height= 6cm]{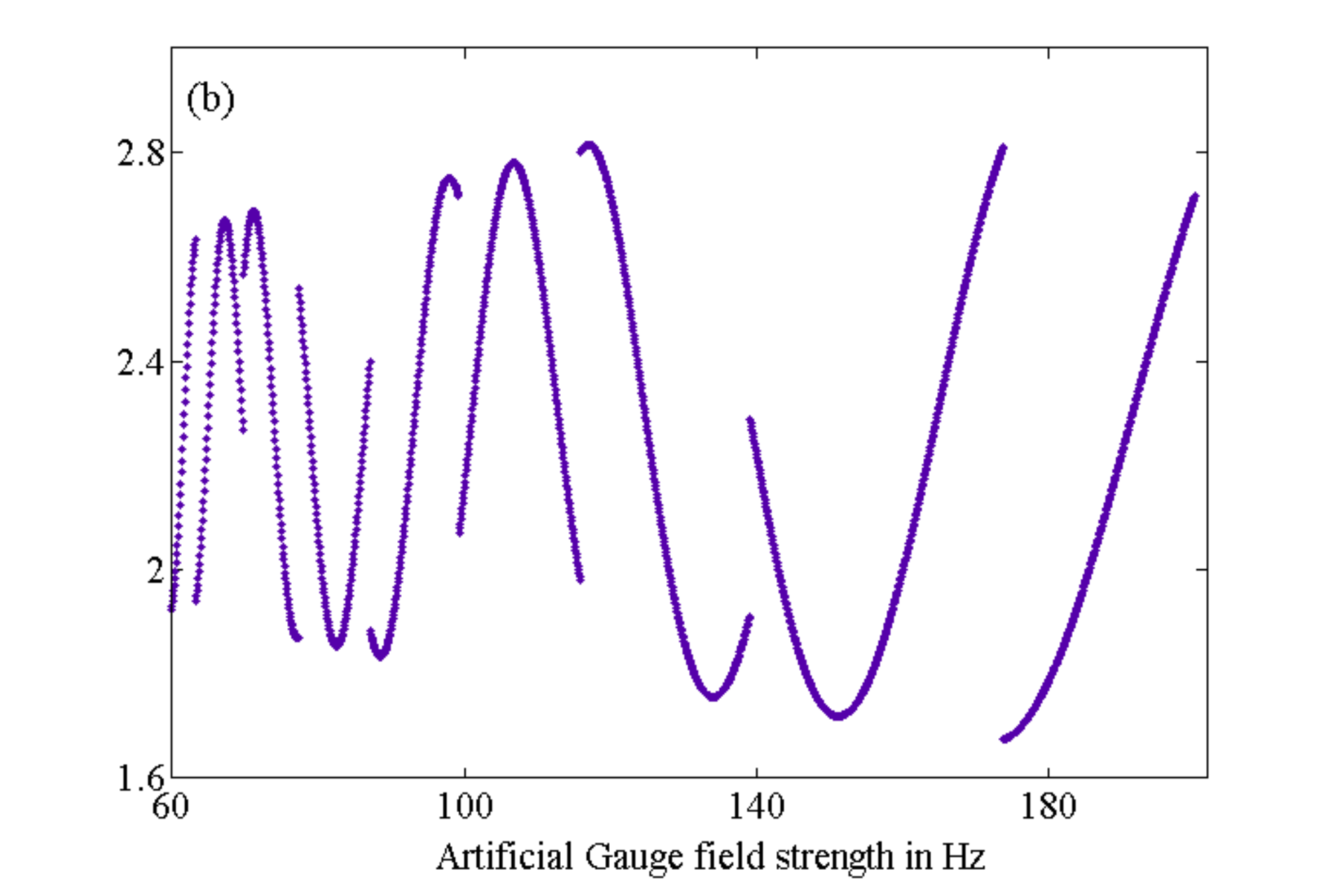}
%\end{subfigure}
\caption{Variation of $S_\nu$ with (a) number of trapped atoms and (b) gauge field strength.}
\label{SB}
\end{figure}
%%%%%%%%%%%%%
The distance between the two turning points of the bistability curve in Fig. \ref{eta}(b) is calculated to be
\beq
h(S_\nu) = \frac{4(\Delta^2 - 3\kappa^2)^{3/2}}{27 S_\nu}. 
\eeq
An  experimentally obtained cavity spectrum
\cite{Eslng, Colombe, Kurn} can be used to extract the corresponding $h(S_\nu)$, and hence the corresponding $S_\nu$,  which can be compared with 
the theoretical value  obtained from Eq. (\ref{Snu}) (Fig. \ref{SB}), provides one the informations about the  Landau levels inside the cavity. 

To summarize we have shown that cavity optomechanics could be very useful tool to explore ultra cold atomic system in a synthetic gauge field, providing us a direct access to the 
atomic Landau levels. Future studies may consider the system in the limit when atom-photon interaction will lead to the reorganization of the many body 
atomic state in presence of the the standing wave in the cavity as well as  similar problem for the bosonic systems, where bosons will prefer to stay in the lowest Landau level. 

One of us (BP) thanks V. Ravishankar for helpful comments.

\newpage
\begin{center}
\textbf{ Supplementary material}
\end{center}
%\maketitle 
%\renewcommand{\thesection}{\Alph{section}}
\numberwithin{equation}{section}

\section{Hamiltonian for synthetic Landau level problem}
There are several established schemes to generate a synthetic magnetic field for charge-neutral ultra cold atoms.
Here we provide some details of the corresponding hamiltonians in these schemes.
In  the scheme developed in NIST \cite{Spielman}, 
a Landau gauge type artificial vector potential was generated by coupling different hyperfine states of the atoms through Raman lasers, which transfers momentum only along the $\hat{x}$ direction. The coupling can vary spatially and leads to  the effective single atom Hamiltonian 
\bea H=H_{1}(k_{x})+[\hbar^{2}(k_{y}^{2}+k_{z}^{2})/2m+V(r)], \nonumber\eea
where \bea H_{1}(k_{x})=\frac{\hbar^{2}\left(k_{x}-\frac{q^{\ast}A_{x}^{\ast}}{\hbar}\right)^{2}}{2m^{\ast}}. \nonumber\eea
Here $A_{x}^{\ast}$ is the engineered vector potential and $q^{*}$ is the fictitious charge.
In this case if the system size is sufficiently large and one considers the bulk of the system,  the hamiltonian resembles  that 
for the charged particle in magnetic field, but with the vector potential in Landau gauge. This hamiltonian is same as the atomic hamiltonian we used 
in the main paper. 

In another set of schemes the trap 
that contains the ultra cold atom is rotated with the help of a moving laser beam \cite{Dalibard} about a particular axis. 
The corresponding problem is considered in the co-rotating frame where the laws of equilibrium thermodynamics 
holds. The corresponding single particle hamiltonian  can be written as
\beq
{H}_L= \frac{|{\bf {p}}|^2}{2M} + \frac{1}{2}M\omega_{xy}^2({x}^2 + {y}^2) + \frac{1}{2}M\omega_z^2z^2 - {\bf \Omega .{r}\times {p}}. \label{Ham1}
\eeq
We assume that the axis of rotation is along $z$ and also the system is very tightly trapped along the $z$-axis, such that it is always in the ground state of the harmonic 
trap potential along the $z$-direction. Under this situation the dynamics along $z$-axis is frozen and the motion is essentially two dimensional and confined to the xy plane such that 
an effective two dimensional hamiltonian can replace \ref{Ham1} ( without the term $\frac{1}{2} M \omega_{z}^{2} z^{2}$) giving  
\bea
 H_{L} & = & \frac{1}{2M}\Big({\bf|p|^2} - 2M{\bf p.\Omega\times r} + |{\bf \Omega\times r}|^2 \Big ) + \frac{1}{2}M\omega_{xy}^2({x}^2 + {y}^2) \nonumber \\
 & = &\frac{1}{2M}{\bf {\Pi}}^2 + \frac{1}{2}M(\omega_{xy}^2 - \Omega^2)({x}^2 + {y}^2),
\label{HLE} \eea
where ${\bf {\Pi}}={\bf {p}} - M{\bf {A}}$ and $ {\bf {A}} = {\bf \Omega \times r}$. For a fast rotating condensate, $\Omega/\omega_{xy}\approx1$ the last term can also be dropped and we obtain Eq. \ref{HLE}. This resembles the hamiltonian for a charged particle in magnetic field written in symmetric gauge and again is  same as the one used as the atomic hamiltonian 
in the main paper. 

The current paper talks about the effect of synthetic magnetic field on a degenerate Fermi gas which is almost non-interacting \cite{Jin}
and we do not consider the case of bosonic or fermionic superfluid.  
It is important to mention here that for such ultra cold atomic superfluid, particularly in  the case of bosons, the effect of rotation 
will create strongly correlated many body vortex state in such system and the observation of Landau level physics under typical experimental
condition is difficult \cite{Fetter} and the NIST scheme is better suited for such system to observe Landau level physics.  
It is not possible to comment at this stage for a trapped atomic system inside a cavity which of these methods will be more suitable to generate synthetic magnetic field. 
However the results we obtained in this paper are independent of choices of synthetic gauge potential. 

\section{Neutral atom in synthetic gauge field}	
\label{Gmn}
In this section we shall elaborate on the Landau level basis that was used to expand the field operators in the main paper. 
The atomic hamiltonian describes an atom moving on the $xy$ plane in the presence of such synthetic guage field along the $z$-direction
An atom in a synthetic field classically  moves in a circular (cyclotron) orbit with angular (cyclotron) frequency $\omega_0$. The velocity is ${\bf v}={\bf \Pi}/M$, and position of the center of cyclotron orbit, called guiding center coordinate is \beq{\bf R}={\bf r}+\frac{\hat{z}\times{\bf v}}{\omega_0}. \eeq

Quantum mechanically the eigenstates and eigenvalues of this Hamiltonian
can be obtained by the usual ladder operator method. The observation that the x and y component kinetic momenta form a canonically conjugate pair allows us to define the ladder operators.
\bea
[\Pi_x, \Pi_y] &=& \frac{-\imath \hbar^2}{l_0^2}. \nn \\
d^\dag &\equiv& \frac{l_0}{\hbar \sqrt{2}}(\Pi_x+\imath \Pi_y),
\eea
These ladder operators obey bosonic commutation relation $[d, d^\dag ]=1$. Hence the Hamiltonian becomes \beq H_L = \hbar \omega_0 (dd^\dag + \frac{1}{2}). \eeq
Because of the fact that the classical kinetic energy is independent of the guiding center coordinate we might expect a degeneracy in the eigenstates with respect to the quantum orbit-center operators $g \equiv \frac{1}{\sqrt{2} l_{0}}(R_x+\imath R_y)$, 
\beq [H_L, g] = 0. \eeq
This commutator can be used to compute the complete spectrum (Landau levels) of the Hamiltonian:
\bea
|n,m\rangle &=& \frac{(d^\dag)^n(g^\dag)^m}{\sqrt{n!m!}}|0,0\rangle, \\
E_{n,m} &=& \hbar \omega_0 (n + \frac{1}{2}). 
\eea
For the choice of symmetric gauge ${\bf A} = \Omega (y,-x,0)$, we obtain
\bea
\langle {\bf r}|0,0\rangle &=& \frac{1}{\sqrt{2\pi l_0^2}} e^{-r^2/4l_0^2}, \nn \\
\langle {\bf r}|n,m\rangle &=& \frac{e^{-|z|^2/4l_0^2}}{\sqrt{2\pi l_0^2}}G_{m+n,n}(iz/l_0), \nn \\
G_{n+m,m}(z) &=& (\frac{m!}{(n+m)!})^{1/2}(- \frac{\imath z}{\sqrt {2}})^nL_m^n (\frac{|z|^2}{2}).
\eea
$L_m^{n}$ is the generalized Laguerre polynomials of degree $m$.  We briefly state some useful properties of the special function $G_{n,n'}$ and refer to  \cite{Macdonald} for more details.

($\mathfrak{a}$) The matrix element of the plane wave operator $e^{-\imath {\bf k.r}}, r=\sqrt{2}l_0(g-d^\dag)$ in the Landau level basis is
\beq
\langle n,m|e^{-\imath {\bf k.r}}|n',m'\rangle = e^{-|l_0k|^2/2}G_{n',n}(l_0k^{*})G_{m',m}(l_0k).
\eeq

($\mathfrak{b}$) Complex conjugates:
\bea
G_{m,m'}(q) &=& G^*_{m,m'}(-q^*) = G^*_{m',m}(-q) = G_{m',m}(q^*) \nn \\
G_{m,m'}(\imath q) &=& G^*_{m,m'}(\imath q^*) = G^*_{m',m}(-\imath q^*) = (-\imath)^{m-m'}G_{m',m}(\imath q^*). \eea

($\mathfrak{c}$) Orthogonality of Landau basis, or that of generalized Laguerre polynomials: 
\beq
\int d^2k e^{-|l_0k|^2/2} G_{m',m}(l_0k)G_{n',n}(l_0k^{*}) = \frac{1}{ 2\pi l_0^2}\delta_{m',n'}\delta_{m,n}.
\eeq

($\mathfrak{d}$) Completeness of Landau basis:

\beq \sum_n G_{m,n}(l_0k_1)G_{n,p}(l_0k_2) = e^{-l_0^2k_1^*k_2/2}G_{m,p}(l_0k_1+l_0k_2) \eeq

($\mathfrak{e}$) The Fourier transform of the product of two functions:

\beq 
\i e^{-\imath {\bf k}.{\bf r}} \langle n',m'|{\bf r}\rangle  \langle{\bf r}|n,m\rangle =  e^{-l_0^2|k|^2/2}G_{n',n}(l_0k^*)G_{m',m}(l_0k),	\label{fourier}  \eeq

\section{Derivation of the effective atom-photon Hamiltonian}\label{Heff}

Using rotating wave and electric dipole approximation, we describe a single atom of this system by the Jaynes-Cummings Hamiltonian \cite{Cummings},\cite{Maschler}.
\bse
\begin{align} 
\hat{H}_L &= \frac{{\bf \hat{\Pi}}^2}{2M} - \hbar \Delta_a \hat{\sigma}^+\hat{\sigma}^-, \\
\hat{H}_C &= -{\hbar{\Delta_c}}\hat{a}^{\dagger}\hat{a}-{\imath}{\hbar}{\eta}(\hat{a}-\hat{a}^\dagger), \\
\hat{H}_I &=-{\imath}{\hbar}g({\bf r})({\hat{\sigma}}^+\hat{a}-{\hat{\sigma}}^-\hat{a}^\dagger).
\end{align}
\ese
Here the cavity mode function is $g({\bf r})=g_0cos({\bf K}.{\bf r})$ and detuning parameters are ${\Delta}_c={\omega}_p-{\omega}_c, {\Delta}_a = {\omega}_p-{\omega}_a$. The atom Hamiltonian contains $\hat{\Pi}$, the kinetic momentum operator. The interaction Hamiltonian includes only the contribution of atom-cavity field interaction. The atomic interaction is dropped because s-wave interaction is not present for fermions due to Pauli's exclusion rule. The atoms can also interact with the pump laser, which coherently drives them. But by choosing a z-axis dependent mode function for this Laser, this interaction term can be dropped. 

We describe this atom-photon many body system using the language of second-quantization. We denote the atomic field operators for annihilating an atom at position ${\bf r}$ in the ground state and excited state by $\Psi_g({\bf r})$ and $\Psi_e({\bf r})$, respectively. They obey usual bosonic commutation relations.

The cavity field operator remains unchanged. Others can be modeled as
\bse \begin{align}
\hat{H}_L&={\i}{\hat{\Psi}}_g^\dag({\bf r})\Big{(}\frac{{\bf \hat{\Pi}}^2}{2M}\Big{)}{\hat{\Psi}}_g({\bf r})+{\i}{\hat{\Psi}}_e^\dag({\bf r})\Big{(}\frac{{\bf \hat{\Pi}}^2}{2M} - \hbar \Delta_a \Big{)}{\hat{\Psi}}_e({\bf r}). \\
\hat{H}_I&= -\imath\hbar\i\hat{\Psi}_g^\dag({\bf r})g({\bf r})\hat{a}^\dag\hat{\Psi}_e({\bf r}) +h.c.
\end{align} \ese
Now we calculate the Heisenberg equation of time evolution for the various operators. 
\bse \begin{align}
\frac{\partial \hat{\Psi}_e({\bf r})}{\partial t} & = \imath[\hat{H}_L + \hat{H}_I, \hat{\Psi}_e({\bf r})]/\hbar = -\imath \Big{(}\frac{{\bf \hat{\Pi}}^2}{2M\hbar} - \Delta_a \Big{)} \hat{\Psi}_e({\bf r})  - g({\bf r})\hat{a} \hat{\Psi}_g({\bf r}),  \\
\frac{\partial \hat{\Psi}_g({\bf r})}{\partial t} & = \imath[\hat{H}_L + \hat{H}_I, \hat{\Psi}_g({\bf r})]/\hbar = -\imath\frac{{\bf \hat{\Pi}}^2}{2M\hbar}\hat{\Psi}_g({\bf r}) + g({\bf r})\hat{a}^\dag\hat{\Psi}_e({\bf r}), 	\label{psigt} \\ 
\frac{\partial \hat{a}}{\partial t} & = \imath[\hat{H}_I + \hat{H}_C, \hat{a}]/\hbar  = \imath \Delta_c \hat{a} +\eta + \i g({\bf r})\hat{\Psi}^\dag_g({\bf r})\hat{\Psi}_e({\bf r}). \label{adt}
\end{align} \ese

At very low temperature there is only weak atomic excitations. And also we can assume $\Delta_a (\sim$100GHz ) $\ll 1/\Gamma (\sim$ 1MHz, $\Gamma$ is natural life time of the atomic excited state), i.e. the atom in its excited state almost instantaneously gets damped to equilibrium, where as $\hat{\Psi}_g({\bf r})$, and $a$ vary in a much slower time scale. In this limit we can adiabatically eliminate the excited states from the dynamics of our system. This also allows us to neglect spontaneous emission, hence avoiding situations like Landau level lasers. Setting $\partial_t \hat{\Psi}_e({\bf r})=0$ we get 
\beq
\hat{\Psi}_e({\bf r}) = -\imath g({\bf r})\hat{a} \hat{\Psi}_g({\bf r})/\Delta_a.
\eeq
Now inserting this expression for $\hat{\Psi}_e({\bf r})$ in equations (\ref{psigt}) we get
\bse \begin{align}
\frac{\partial \hat{\Psi}_g({\bf r})}{\partial t} & = -\imath\frac{{\bf \hat{\Pi}}^2}{2M\hbar}\hat{\Psi}_g({\bf r}) - \imath g^2({\bf r})\hat{a}^\dag \hat{a}\hat{\Psi_g}({\bf r})/\Delta_a, \\
\frac{\partial \hat{a}}{\partial t} & = \imath \Big{[}\Delta_c - \frac{1}{\Delta_a}\i g^2({\bf r})\hat{\Psi}^\dag_g({\bf r})\hat{\Psi}_g({\bf r})\Big{]}\hat{a} +\eta. 
\end{align} \ese
Now the effective Hamiltonian $\hat{H}_{eff}$ that will receover the above dynamics must obey the same Heisenber operator equation, namely 
\bea \frac{\partial \hat{\Psi}_g({\bf r})}{\partial t} = \imath[\hat{H}_{eff}, \hat{\Psi}_g({\bf r})]/\hbar, \nn \\
\frac{\partial \hat{a}}{\partial t} = \imath[\hat{H}_{eff}, \hat{a}]/\hbar.
\eea  
This gives the effective hamiltonian
\beq
\hat{H}_{eff} = \i \Pds \Big{[}{\bf \Pi}^2/2M + \hbar U_0cos^2({\bf K}.{\bf r})\hat{a}^\dagger \hat{a}\Big{]}{\Ps} - {\hbar{\Delta_c}}\hat{a}^\dag \hat{a}-\imath \hbar \eta (\hat{a}-\hat{a}^\dagger).
\eeq
Where $U_0$ is the coupling amplitude =$g_0^2/({\omega}_L-{\omega}_a)$, $g_0$ is the single photon Rabi frequency. As we have already eliminated the $\hat{\Psi}_e({\bf r})$, for notational simplifications we replace $\hat{\Psi}_g({\bf r})$ by ${\Ps}$. It can be checked  that if we first eliminate the excited state in the single particle dynamics and subsequently go to the field theoretic considerations, then we get the same effective Hamiltonian.

\section{Derivation of the Bosonized hamiltonian given in Eq. (14)}
In this section we shall provide a derivation of the bosonized atom-photon hamiltian given in Eq. (14) 
of the main paper. 
To this purpose we shall first derive the non-interacting hamiltonian given in Eq. (12)

We calculate the commutation relation between $\hat{H}_L$ and the bosonic creation operator $\hat{b}_n^\dag$ 
\beq
[\hat{H}_L, \hat{b}^\dag_p({\bf k})] = \frac{\hbar {\omega}_0 e^{-(l_0|k|)^2/4}}{\sqrt{pN_\phi J_p^2(k R_\nu)}} \sum\limits_{q,m,m'=0}^{\infty}\sum\limits_{n,s=0}^{\infty} (n+\frac {1}{2}) G_{q+p,q}(l_0 k^*)G_{m',m}(l_0k) [\hat{c}_{n,s}^{\dagger}\hat{c}_{n,s}, \hat{c}^\dag_{q+p,m'}\hat{c}_{q,m}], \nn 
\eeq
Now, 
\bea
[\hat{c}_{n,s}^{\dagger}\hat{c}_{n,s}, \hat{c}^\dag_{q+p,m'}\hat{c}_{q,m}] &=& \hat{c}_{n,s}^{\dagger}\hat{c}_{n,s}\hat{c}^\dag_{q+p,m'}\hat{c}_{q,m} -\hat{c}^\dag_{q+p,m'}\hat{c}_{q,m} \nn \\
&=& \hat{c}_{n,s}^{\dagger}\{\hat{c}_{n,s},\hat{c}^\dag_{q+p,m'}\}\hat{c}_{q,m} - \hat{c}^\dag_{q+p,m'}\{\hat{c}_{q,m},\hat{c}_{n,s}^{\dagger}\}\hat{c}_{n,s} \nn \\
&=& \hat{c}_{n,s}^{\dagger}\delta_{n,q+p},\delta_{s,m'}\hat{c}_{q,m} - \hat{c}^\dag_{q+p,m'}\delta_{n,q}\delta_{s,m}\hat{c}_{n,s}. \nn 
\eea
In coming from first to second step of the above commutation we have used the anti-commutation relation and cancelled the terms $\hat{c}^\dag_{q+p,m'}\hat{c}_{n,s}^{\dagger}\hat{c}_{q,m}\hat{c}_{n,s}$, and $\hat{c}_{n,s}^{\dagger}\hat{c}^\dag_{q+p,m'}\hat{c}_{n,s}\hat{c}_{q,m}.$
Now effecting the summations over $n, s$, we get
\bea
[\hat{H}_L, \hat{b}^\dag_p({\bf k})] &=& \frac{\hbar {\omega}_0 e^{-(l_0|k|)^2/4}}{\sqrt{pN_\phi J_p^2(k R_\nu)}} \sum\limits_{q,m,m'=0}^{\infty}G_{q+p,q}(l_0 k^*)G_{m',m}(l_0k)\nn \\ & & \Big((p+q+\frac {1}{2}) \hat{c}_{q+p,m'}^{\dagger}\hat{c}_{q,m}-(q+\frac{1}{2})\hat{c}^\dag_{q+p,m'}\hat{c}_{q,m}\Big) \nn \\
&=& \hbar \omega_0 p \hat{b}^\dag_p({\bf k}) \nn
\eea
Similarly we obtain $[\hat{H}_L, \hat{b}_p({\bf k})] = \hbar \omega_0 p \hat{b}_p({\bf k}).$ Hence, an effective Hamiltonian that can describe the dynamcis of the system (i.e. which also obeys the above commutation relations) can be obtained as
\beq 
\hat{H}_L= \hbar \sum\limits_{p=1}^{\infty} \sum\limits_{\bf k} p\omega_0  b_p^\dagger({\bf k})b_p({\bf k}). \eeq

To derive the bosonized  atom-photon interaction  given in Eq. (13) we do the following. Expanding the atom-photon interaction Hamiltonian in the Landau basis. Recognizing $4cos^2({\bf K.r})=2+2cos(2{\bf K.r})= 2+e^{-\imath 2{\bf K.r}}+e^{\imath 2{\bf K.r}}$ we write 
\bea
\hat{H}_I &=& \i \Pds[\hbar U_0cos^2({\bf K.r})a^{\dagger}a]\Ps \nonumber \\
& = & \frac{\hbar U_0}{4} \Big{\{}2\i \Pds \Ps + \i \Pds e^{-\imath 2{\bf K.r}}\Ps + \i \Pds e^{\imath 2{\bf K.r}}\Ps \Big{\}} a^{\dagger}a. \nn
\eea
%Using expansion of $\Ps, \Pds$, and the property of fourier transform of the product of two functions we get 
%\beq 
%\i e^{-\imath {\bf k}.{\bf r}} \langle n',m'|{\bf r}\rangle  \langle{\bf r}|n,m\rangle =  e^{-2l_0^2|k|^2}G_{n',n}(2l_0k^*)G_{m',m}(2l_0k),	\nn \eeq

% NOTE: The formula II.12 will be modified to (because of the typo in Girvin's paper) 
We also have \cite{Macdonald}
\beq 
\i e^{-\imath {\bf k}.{\bf r}} \langle n',m'|{\bf r}\rangle  \langle{\bf r}|n,m\rangle =  e^{-l_0^2|k|^2/2}G_{n',n}(l_0k^*)G_{m',m}(l_0k),	\label{fourier}  \eeq
and 
%also II.8
\beq
\langle n,m|e^{-\imath {\bf k.r}}|n',m'\rangle = e^{-|l_0k|^2/2}G_{n',n}(l_0k^*)G_{m',m}(l_0k). 
\eeq
With these we get, 
\bea
\i \Pds \Ps & = & \i e^{-\imath 0.{\bf r}}\Pds \Ps = \sum\limits \cdp \c \i e^{-\imath 0.{\bf r}} \langle n',m'|{\bf r}\rangle  \langle{\bf r}|n,m\rangle  \nn \\
& =& \sum\limits \hat{c}_{n',m'} \c G_{n',n}(0) G_{m',m}(0) = \sum\limits \hat{c}_{n',m'} \c \delta_{n,n'}\delta_{m',m} = \sum\limits \cd \c  \nn \\
&=& N, \nn	\eea
\bea
\i \Pds e^{-\imath 2{\bf K.r}}\Ps &=& \sum\limits \hat{c}_{n',m'} \c \i e^{-\imath 2{\bf K.r}} \langle n',m'|{\bf r}\rangle \langle{\bf r}|n,m\rangle  \nn \\
& =& \sum\limits \hat{c}_{n',m'} \c  e^{-2l_0^2|K|^2}G_{n',n}(2l_0K^*)G_{m',m}(2l_0K), \nn
\eea
\bea
\i \Pds e^{\imath 2{\bf K.r}}\Ps &=& \sum\limits \hat{c}_{n',m'} \c \i e^{\imath 2{\bf K.r}} \langle n',m'|{\bf r}\rangle \langle{\bf r}|n,m\rangle  \nn \\
& =& \sum\limits \hat{c}_{n',m'} \c  e^{-2l_0^2|K|^2}G_{n',n}(-2l_0K^*)G_{m',m}(-2l_0K) . \nn
\eea
Here the summation is performed for all of $n,n',m,m'$, and the full expression for the Hamiltonian becomes 
\bea
\hat{H}_I&=&\frac{\hbar U_0}{4}\Big{\{} N + \sum\limits \hat{c}_{n',m'} \c e^{-2(|K|l_0)^2} \Big (G_{n',n}(2 K^*l_0)G_{m',m}(2Kl_0) + \nonumber \\ 
& & G_{n',n}(-2K^*l_0)G_{m',m}(-2Kl_0)\Big )\Big{\}}\hat{a}^\dagger \hat{a}  \label{HFermi}
\eea 

For bosonizing the interacting Hamiltonian
\bea
[\hat{H}_I,\hat{b}^\dag_q({\bf k})] &=& \sqrt{N_\phi qJ_q^2(2k R_\nu)}\frac{\hbar U_0e^{-2(|K|l_0)^2}e^{-(|k|l_0)^2/2}}{4} \sum\limits_{n,n'=0}^{\infty}\sum\limits_{m,m'=0}^\infty \sum\limits_{l,r,r'=0}^\infty [\hat{c}^\dag_{n',m'}\hat{c}_{n,m} , \hat{c}^\dag_{l+q,r'}\hat{c}_{l,r} ] \hat{a}^\dagger \hat{a} \nn \\
& & \Big (G_{n',n}(2 K^*l_0)G_{m',m}(2Kl_0)G_{l+q,l}(l_0 k^*)G_{r',r}(l_0k) \nonumber \\ 
& & + G_{n',n}(-2K^*l_0)G_{m',m}(-2Kl_0)G_{l+q,l}(l_0 k^*)G_{r',r}(l_0k) \Big ) 
\eea
By substituting $n'$ to $n+p$ the first part of the sum in the above equation becomes comparable to Eq. (9) hence vanishes. And the second part being comparable with Eq. (7) gives rise to the delta function,
\bea
[\hat{H}_I, \hat{b}_q^{\dagger}({\bf k})] 
&=& \frac{\hbar U_0}{4} \sqrt{N_\phi qJ_q^2(2K R_\nu)} \delta({\bf k}-2{\bf K}) \hat{a}^\dagger \hat{a}.
\eea
Similarly we can show 
\beq
[\hat{H}_I, \hat{b}_q({\bf k})]=\frac{\hbar U_0}{4} \sqrt{N_\phi qJ_q^2(2K R_\nu)} \delta({\bf k}-2{\bf K}) \hat{a}^\dagger \hat{a}.
\eeq
This is the bosonized atom-photon interaction given in the main paper 

Therefore, an effective Hamiltonian describing such dynamics is found to be
\beq
\hat{H}_I = \frac{\hbar U_0}{4}\sum\limits_{p=1}^{\infty} \sqrt{N_\phi pJ_p^2(2K R_\nu)}\Big (\hat{b}_p^\dagger(2{\bf K}) +\hat{b}_p(2{\bf K})\Big ) \hat{a}^\dagger \hat{a}
\eeq

Adding these two hamiltonians namely $\hat{H}_{I}$ and $\hat{H}_{L}$, we get the effective hamiltonian (14) in the main paper.

\section{Derivation of Heisenberg equation of motion} \label{Heisenberg}
Here we provide the derivations for the Heisenberg equations for the bosonic operators $\hat{b}_p, \hat{b}_p^\dag$, used in the main text.
The cavity field part of the Hamiltonian commute with $\hat{b}_p, \hat{b}_p^\dag$. Now, 
\bea
[\hat{H}_{eff},\hat{b}_p(2{\bf K})] &=& \hbar \sum\limits_{p'=1}^{\infty} \Big \{ \sum\limits_{\bf k} {p'}\omega_0 [\hat{b}_{p'}^\dagger({\bf k})\hat{b}_{p'}({\bf k}), \hat{b}_p(2{\bf K})] + \delta_{p'} \sqrt{p'}[\Big (\hat{b}_{p'}^\dagger(2{\bf K})+\hat{b}_{p'}(2{\bf K})\Big ),\hat{b}_p(2{\bf K})]\hat{a}^\dagger \hat{a} \Big \}, \nn \\
&=& - \hbar p\omega_0 \hat{b}_p({2\bf K}) -\hbar \delta^\nu_p \sqrt{p} \hat{a}^\dagger \hat{a}. \nn \label{bp} 
\eea
Using the  hermiticity  we get
 \beq
[\hat{H}_{eff},\hat{b}_p^\dag(2{\bf K})]= -[\hat{H}_{eff},\hat{b}_p(2{\bf K})]^\dag = \hbar p\omega_0 \hat{b}_p^\dag({2\bf K}) + \hbar \delta^\nu_p \sqrt{p} \hat{a}^\dagger \hat{a}. \label{bpd} \nn \eeq

By using the above results we derive the Heisenberg equations for the quadratures 
\bea
\frac{d\hat{X}_L}{dt} &=&\frac{\imath}{\hbar}[\hat{H}_{eff},\hat{X}_L] =\frac{\imath}{\hbar \sqrt 2}([\hat{H}_{eff},\hat{b}_p^\dagger(2{\bf K})] + [\hat{H}_{eff},\hat{b}_p(2{\bf K})])  \nn \label{XLt} \\
&=&p\omega_0 \hat{P}_L, \\
\frac{d\hat{P}_L}{dt} &=&\frac{\imath}{\hbar}[\hat{H}_{eff},\hat{P}_L] =\frac{-1}{\hbar \sqrt 2}([\hat{H}_{eff},\hat{b}_p^\dagger(2{\bf K})] - [\hat{H}_{eff},\hat{b}_p(2{\bf K})]) \nn \\
&=& -p\omega_0\hat{X}_L - \delta^\nu_p \sqrt{2p} \hat{a}^\dagger \hat{a}.  \label{Pt} 
\eea
In order to derive the Heisenberg equation for field quadratures we first calculate the commutations, $[\hat{H}_{eff},\hat{a}], [\hat{H}_{eff},\hat{a}^\dag] $. We use the relation $[\hat{a},\hat{a}^\dag]=1.$
\bea
[\hat{H}_{eff},\hat{a}] &=& \hbar \sum\limits_{p=1}^{\infty} \Big \{ \delta^\nu_p \sqrt{p}\Big (\hat{b}_p^\dagger(2{\bf K})+\hat{b}_p(2{\bf K})\Big )[\hat{a}^\dag \hat{a},\hat{a}]\Big \} +\hbar \Delta [\hat{a}^\dag \hat{a},\hat{a}] + \imath\hbar\eta[\hat{a}^\dag,\hat{a}] \nn \\
& = & -\hbar \sum\limits_{p=1}^{\infty} \Big \{ \delta^\nu_p \sqrt{p}\Big (\hat{b}_p^\dagger(2{\bf K})+\hat{b}_p(2{\bf K})\Big )\hat{a} \Big \} -\hbar \Delta \hat{a} - \imath\hbar\eta \nn .
\eea

The equation of evolution of the field operators is
\bea
\frac{d\hat{a}}{dt} &= \frac{\imath}{\hbar}[\hat{H}_{eff},\hat{a}] = -\imath \sum\limits_{p=1}^{\infty} \delta^\nu_p \sqrt{2p}\hat{X}_L \hat{a} - \imath\Delta \hat{a} + \eta - \kappa \hat{a} + \sqrt{2\kappa}\hat{a}_{in}  \label{adt}
\eea 

\section{Linear stability analysis from Heisenberg equations under restricted condition: Routh-Hurwitz stability criterion}
\label{Routh}
In the main draft we already pointed out the difficulties in doing a stability analysis using Heisenberg equation of motion.
Nevertheless, in this section, we provide a  linear stability analysis around  the steady state condition for intercavity 
photon number from the Heiseberg equation of motion under a very restrictive situation.
We define such a situation, by considering when only transition from  the filled Landau level to a single unfilled Landau level is allowed and all other transitions are supressed.
Under this restricted situation 
$\mathcal{O}(t) = [\hat{X}_L (t), \hat{P}_L (t) , \hat{X} (t) , \hat{P} (t) ]^T,$  for a single $p$,  
with $\hat{X} =  (\hat{a}^\dag + \hat{a})/\sqrt2, \hat{P}= \imath(\hat{a}^\dag - \hat{a})/\sqrt2$ being cavity quadratures. From the Heisenberg equation defined in the main text 
after setting $\mathcal{O}(t) = \mathcal{O}^{(s)} + \delta \mathcal{O}(t)$, where the superscript $^{s}$ stands for steady state condition, we get

\bea
&\mathcal{\delta O'}(t)& = \qquad \qquad \qquad \qquad J \qquad \qquad \qquad \qquad . \mathcal{\delta O}(t) \nn \\
&\bem
      \delta \hat{X}'_L \\
      \delta \hat{P}'_L \\
      \delta \hat{X}' \\
	\delta \hat{P}' \\
  \eem &
= 
	\bem
	0 & p\omega_0 & 0 & 0 \\
	-p\omega_0 & 0 & -2\delta^\nu_p \sqrt{p}a_s & 0 \\
	0 & 0 & -\kappa & \Delta' \\
	-\sum\limits_{p=1}^{\infty} 2\delta^\nu_p \sqrt{p}a_s & 0 & -\Delta' & -\kappa \\
    \eem . 
 	\bem  
      \delta \hat{X}_L \\
      \delta \hat{P}_L \\
      \delta \hat{X} \\
	\delta \hat{P} \\
\eem \label{J}
\eea
Here $\Delta' = \Delta - S_\nu n_{ph}.$ 

The dynamical behavior is now given by $\mathcal{O}(t) = e^{\lambda t}\mathcal{O}^{(s)}$, where $\lambda$ is the eigenvalue of the  drift matrix $J$.
The stationary state is stable iff and only if all the roots of the characteristic equation have a negative real part. We use the Routh-Hurwitz theorem. 
The imaginary part of eigenvalue corresponds to the oscillation frequency of each quadratures. 
At the crest of the bistability curve, the imaginary part becomes zero,  confirming a bistable behavior of the steady state.
Relevant to our case, we present the analysis \cite{Marden} for a 4 degree polynomial equation, which can very easily be generalized to equation of any finite order. Consider
\beq
\Lambda(\lambda) = \lambda^4 + a_1\lambda^3 + a_2\lambda^2 + a_3\lambda + a_4 = 0 \label{charac}, 
\eeq
According to the  extended Routh-Hurwitz theorem we introduce the ordered sequence $ S = (1, D_1, D_2/D_1, D_3/D_2, a_4)$ where

\bea
D_1&=& a_1. \nn \\
D_2 &= &\begin{vmatrix} a_1 & 1 \\ a_3 & a_2 \end{vmatrix} = a_1a_2-a_3, \nn \\
D_3 &=&\begin{vmatrix} a_1 & 1 & 0 \\ a_3 & a_2 & a_1 \\ 0 & a_4 & a_3 \end{vmatrix}=a_1a_2a_3-a_1^2a_4-a_3^2 .
\eea

The theorem reads the number of roots of $\Lambda(\lambda)$ having \textit{positive real} parts is equal to the number of \textit{changes of signs} in the ordered sequence $S$. So demanding all the members of the sequence to be positive (that is the sign of the first member '1') we can extract the stability criterion of the system. 

Now we apply this theorem to our system. The eigenvalues of the  $ 4 \times 4$ drift matrix given in the main paper posses characteristic equation of the same form as 
(\ref{charac}), where
\begin{subequations}
\begin{align}
a_1 &= 2\kappa,\\
a_2 &= \kappa^2 + \Delta'^2 + p^2\omega_0^2,\\
a_3 &= 2p^2\omega_0^2\kappa,\\
a_4 &= p^2\omega_0^2(\kappa^2 + \Delta'^2) - 4pn_{ph}\delta^\nu_p\sqrt{p}\sum \limits_{p=1}^{\infty} \delta^\nu_p\sqrt{p}\omega_0\Delta'. \end{align}
\end{subequations}
So the elements of the orderes sequence are
\bea
D_1 &=& 2\kappa , \nonumber \\ 
D_2 &=& \kappa^2 + \Delta'^2 , \nonumber \\
D_3 &=& 16pn_{ph}\kappa^2\delta_p\sqrt{p}\sum \limits_{p=1}^{\infty} \delta^\nu_p\sqrt{p}\omega_0\Delta', \nonumber \\
a_4 &=& p^2\omega_0^2(\kappa^2 + \Delta'^2) - 4pn_{ph}\delta^\nu_p\sqrt{p}\sum \limits_{p=1}^{\infty} \delta^\nu_p\sqrt{p}\omega_0\Delta'.
\eea
We observe $D_1, D_2, D_3$ are always positive. And $a_4 \geq 0$ gives us the stability criterion for the system, i.e. 
\beq
n_{ph} \leq \frac{\sqrt {p}\omega_0(\kappa^2 + \Delta'^2)}{4\Delta' \delta^\nu_p\sum \limits_{p=1}^{\infty} \delta^\nu_p\sqrt{p}}.
\label{V5}\eeq

We can rewrite Eq. \ref{V5}  by introducing $C_p = \frac{\sqrt{p}}{4\delta_p^\nu \sum_{p=1}^{\infty}\delta_p^\nu \sqrt{p}},$ we get
\bea
n_{ph} &\leq& C_p \frac{\kappa^2 + \Delta'^2}{\Delta'}. \nn \\
\text{Or, } n_{ph}(\Delta - S_\nu n_{ph}) &\leq& C_p (\kappa^2 + S_\nu^2 n^2_{ph} + \Delta^2 - 2S_\nu n_{ph}\Delta). \nn \\
\text{Or, } \qquad \qquad \qquad \ 0 &\leq& (C_pS_\nu^2 + S_\nu) n^2_{ph} - (2C_pS_\nu \Delta + \Delta) n_{ph} + C_p (\kappa^2+\Delta^2) . 
\eea
Since $C_pS_\nu^2 + S_\nu$ is always positive, the above inequation can hold only when $n_{ph} \ \epsilon \ \mathfrak{R}/(y_-,y_+), $ i.e. the set of real numbers $\mathfrak{R}$, excluding the open interval between the two roots ($y_-$ and $y_+$) of the above quadratic expression in $n_{ph}$, 
\bea
y_\pm = \frac{2C_pS_\nu \Delta + \Delta \pm (\Delta^2 - 4C_pS_\nu\kappa - 4 C_p^2S_\nu^2\kappa^2 \ )^{1/2} }{C_pS_\nu^2 + S_\nu}. 
\eea
The coefficients of the above equation involves a single $p$ as expected. 
Therefore by applying  R-H theorem under such restricted condition for single $p$, we find  that when $n_{ph}$ belongs to $(y_-,y_+)$ the system is unstable, otherwise stable.
The range of instability here of course does not correspond to a practical situation. But the  existence of two bounds and the quadratic form of the stability equation is similar to the 
corresponding situation in Eq. (20) in the main draft.

%\subsection{Issue-II: Stability}

%\textit{The results are unchanged but few things we've misunderstood. I request you to go through the paragraphs near Eq. 4.3-4.11 in Lugiato's paper. What's indended from this analysis is- we %should get that the negative slope region in $n_{ph} vs \eta$ plot is unstable. So if you take derivative of the Eq. \ref{EOS} by treating $n_{ph}$ as y and $\eta^2$ as x and impose $dy/dx \geq 0$ %you'll get $3S_\nu^2 - 4S_\nu\Delta y + (\kappa^2+\Delta^2) \geq 0$. This is what should Eq. D5 look like.}

\end{document}